# Hexagonal High-Entropy Alloys

Michael Feuerbacher*, Markus Heidelmann and Carsten Thomas

*Peter Grünberg Institut PGI-5, Forschungszentrum Jülich GmbH, D-52425 Jülich, Germany*

We report on the discovery of a high-entropy alloy with a hexagonal crystal structure. Equiatomic samples in the alloy system Ho-Dy-Y-Gd-Tb were found to solidify as homogeneous single-phase high-entropy alloys. The results of our electron diffraction investigations and high-resolution scanning transmission electron microscopy are consistent with a Mg-type hexagonal structure. The possibility of hexagonal high-entropy alloys in other alloy systems is discussed.

**Keywords:** High-Entropy Alloy, Structure, Transmission Electron Microscopy

High-entropy alloys (HEAs) constitute a novel field in materials science, currently attracting increasing research interest. The basic idea, pioneered by Yeh in 2004, [1] is to study alloys composed of multiple principal elements solidifying as metallic solid solutions on a simple crystal lattice. Usually, alloys with 4 to 9, occasionally up to 20 components [2] are considered.

In a multicomponent alloy, the total free energy is largely determined by the free energy of mixing, $\Delta G_{mix} = \Delta H_{mix} - T\Delta S_{mix}$, where $\Delta H_{mix}$ is the mixing enthalpy and $\Delta S_{mix}$ the mixing entropy. In equimolar or near equimolar compositions, the mixing entropy can become dominant, stabilizing a random solid solution, if the elements are chosen such that the mixing enthalpy is neither too high positive (leading to segregation) nor too high negative (leading to the formation of ordered structures). In the ideal case the resulting material is a perfectly disordered solid solution with a simple average crystal structure. HEAs are thus related to metallic glasses, but in contrast to the latter, they are thermodynamically stable, in particular at high temperatures.

Up to now numerous alloy systems have been reported in which HEAs exist. Without exception only HEAs with body-centered or face centered cubic structures have been found. In this paper, we report on the observation of a HEA with a hexagonal crystal structure. We have prepared samples of equiatomic composition in the rare earth (RE) system Ho-Dy-Y-Gd-Tb and characterized them by means of transmission electron microscopy (TEM), scanning electron microscopy (SEM), energy dispersive x-ray diffraction (EDX), and scanning transmission electron microscopy (STEM).

---

* Corresponding author. Email: m.feuerbacher@fz-juelich.de

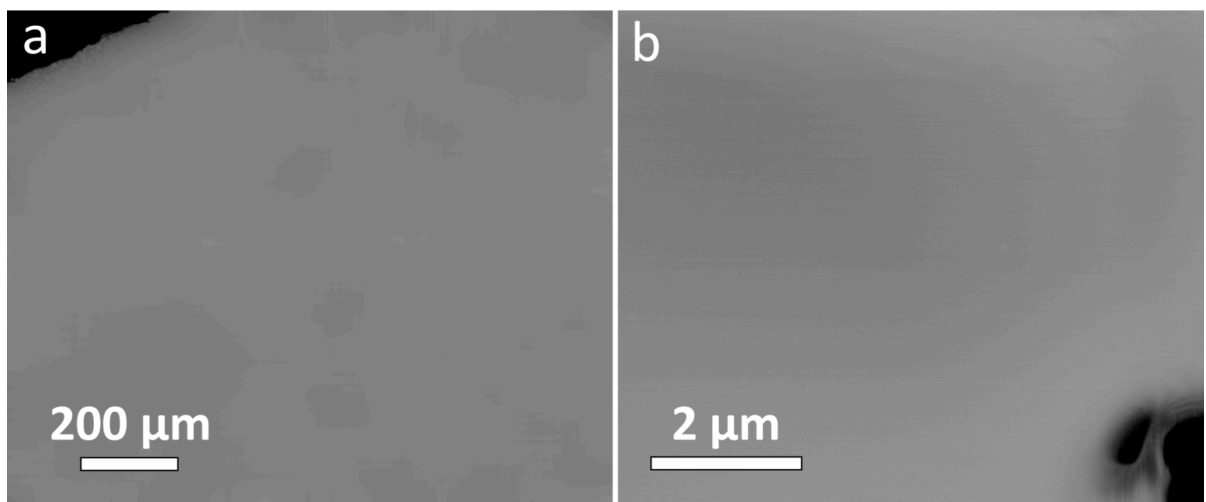

Figure 1. (a) SEM micrograph using a backscattered-electron detector and (b) HAADF-STEM overview of a FIB sample of equiatomic Ho-Dy-Y-Gd-Tb. The material is perfectly homogeneous; no features due to composition variation, precipation, dendrite formation etc. can be seen. In both micrographs sample edges (dark) are visible.

We have prepared ingots of equiatomic Ho-Dy-Y-Gd-Tb in a high-frequency levitation furnace under 1 bar Ar atmosphere using 3N high-purity elements. Each ingot was remelted four times to ensure to achieve proper homogenization of the sample. Specimens for microstructural characterization were prepared by means of a dual-beam FEI Helios Nanolab 400S focused ion-beam (FIB) from the centres of the ingots. TEM was carried out using a Philips CM20 transmission electron microscope operated at 200 kV, SEM work using a JEOL 840 equipped with an EDAX energy dispersive analytical system, and STEM work using a probe corrected FEI Titan 80 – 300 operated at 300 kV and equipped with a high-angle annular dark-field (HAADF) detector.

Fig. 1 (a) displays a scanning electron micrograph of the material using a composition-sensitive backscattered-electron detector. Fig. 1(b) displays a STEM micrograph of the thin part of a typical FIB sample taken using a Z-contrast sensitive HAADF detector. Both images demonstrate that the sample is perfectly homogeneous, in particular there are no composition variations, no precipitates or secondary phases and no dendrites. We obtain these results consistently with FIB samples taken from various positions in the ingots. Note also that the cooling rate after levitation melting is sufficiently slow to allow for the formation of a homogeneous single phase. The average composition measured by EDX at eight different specimen positions is Ho 19.8 at.%, Dy 20.0 at.%, Y 20.4 at.%, Gd 20.1 at.%, Tb 19.7 at.%. Within the experimental uncertainty this corresponds to the nominal equiatomic sample composition, and hence we conclude that the material solidifies congruently.

Fig. 2(a) is an electron diffraction pattern taken at the [0 0 0 1] zone axis. The pattern displays perfect hexagonal symmetry, a number of reflections are indexed.

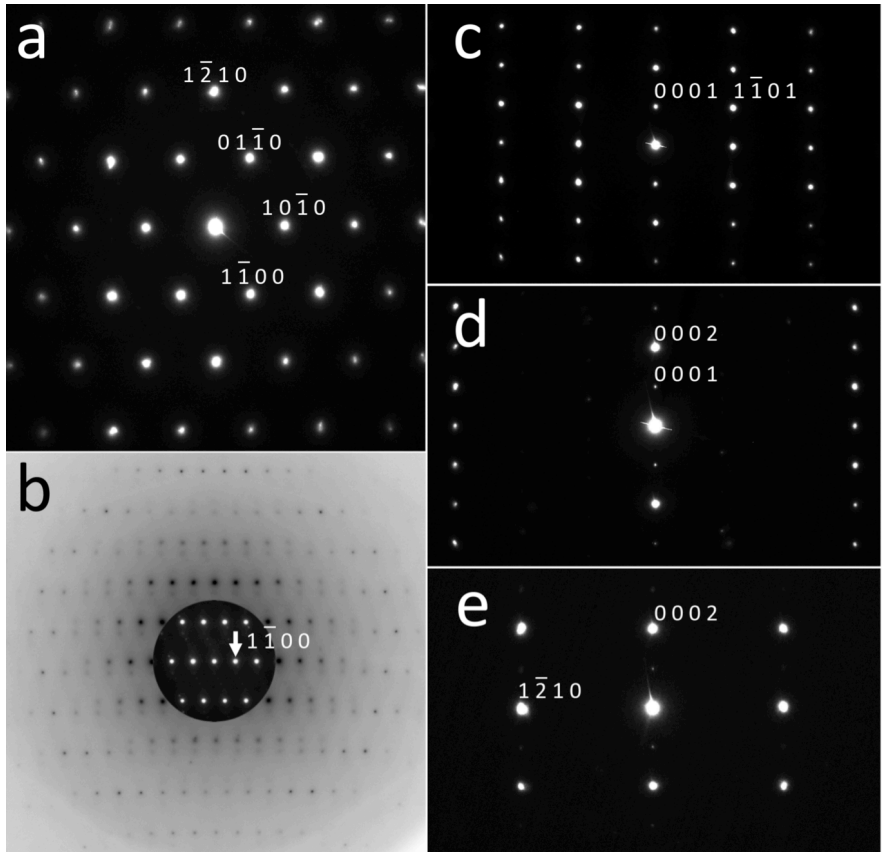

Figure 2. Electron diffraction patterns along different zone axes of the hexagonal structure: (a) [0 0 0 1] zone axis. (b) [1 1 $\bar{2}$ 1] zone axis: overlay of a short-time-exposure displaying zero-order Laue zone reflections and a long-time-exposure (inverted to enhance the weak spots) including first-order Laue zone reflections. (c) [1 1 $\bar{2}$ 0] zone axis. (d) [4 1 $\bar{5}$ 0] zone axis. (e) [1 0 $\bar{1}$ 0] zone axis.

In order to determine the space group we follow a procedure described in [3]. Fig. 2 (b) is an overlay of two electron diffraction patterns at the [1 1 $\bar{2}$ 1] zone axis, a short-time-exposure to display zero-order Laue zone reflections and a long-time-exposure (inverted to enhance the weak spots) including first-order Laue zone reflections. The whole pattern has an *m* symmetry, which corresponds to the point group 6/*mmm* in a hexagonal system. Figs. 2 (c) to (e) show diffraction patterns at the [1 1 $\bar{2}$ 0], [4 1 $\bar{5}$ 0], and [1 0 $\bar{1}$ 0] zone axes, respectively, which are all perpendicular to [0 0 0 1].

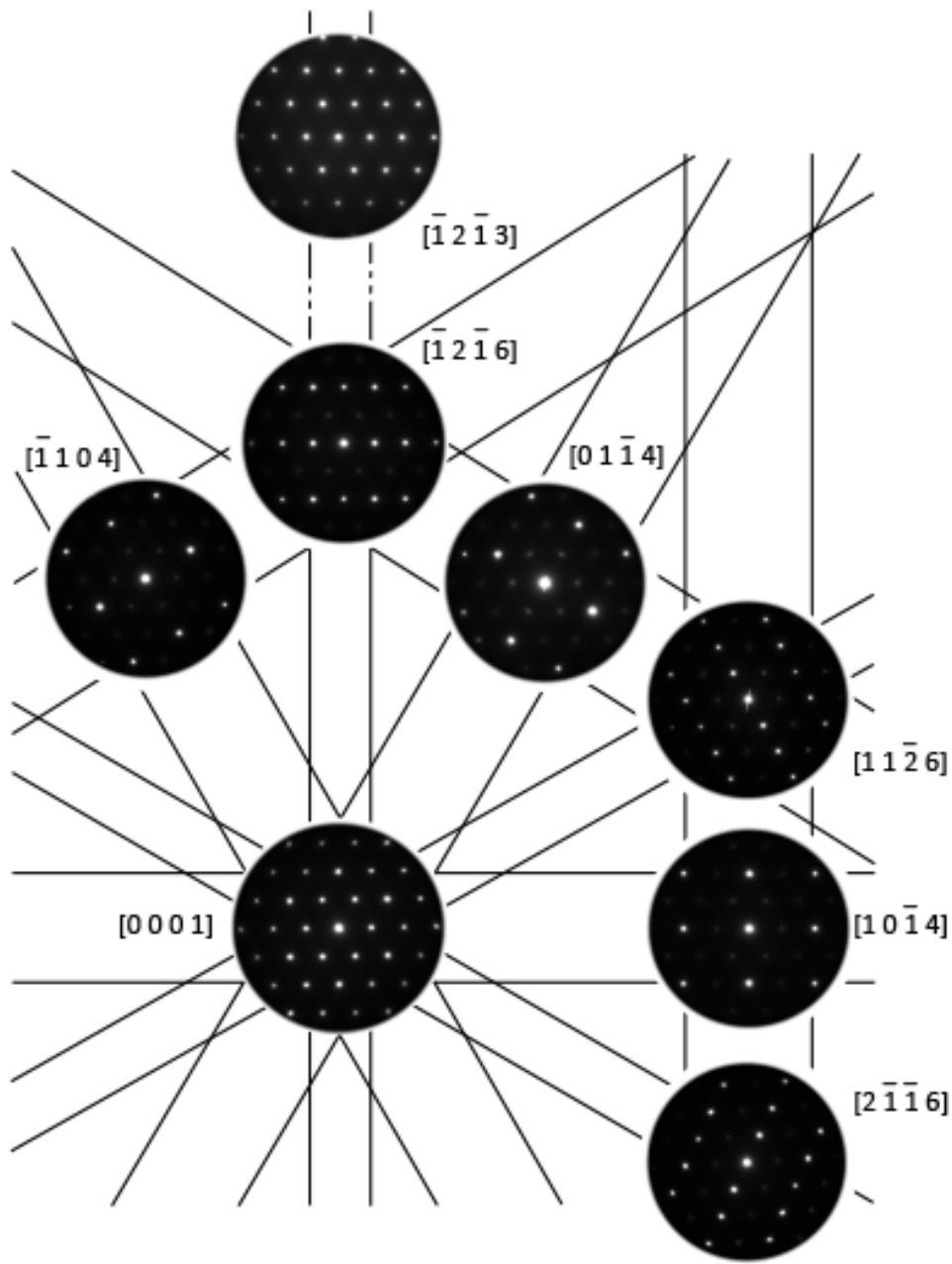

Figure 3. Arrangement of zone axes and corresponding electron diffraction patterns, indexed according to a hexagonal Mg-type structure. The lines correspond to Kikuchi lines as seen in the TEM. Only half of the patterns surrounding [0 0 0 1] are shown. The indicated angle between [$\bar{1}$ 2 $\bar{1}$ 3] and [$\bar{1}$ 2 $\bar{1}$ 6] is not to scale as indicated by the broken lines.

The (0 0 0 1) reflection is visible at the [1 1 $\bar{2}$ 0] zone axis but diminishes when tilting to [4 1 $\bar{5}$ 0] and essentially vanishes when tilting further to [1 0 $\bar{1}$ 0]. This indicates the presence of a $6_3$ screw axis parallel to [0 0 0 1]. In the [1 1 $\bar{2}$ 0] pattern the (1 $\bar{1}$ 0 1) reflections are present, which rules out the P$6_3$/*mcm* case [4]. Consequently the space group must be P$6_3$/*mmc*. Under stern scrutiny, an extremely weak residual intensity of the (0 0 0 1) reflection in the [1 0 $\bar{1}$ 0] zone can be seen. This may be due to a minor deviation from perfect disorder and is neglected in our analysis.

| Zone | Angle to [0 0 0 1] exp./deg | Angle to [0 0 0 1] calc./deg | $R_{hkil}$ ratio exp. | Inverse $d_{hkil}$ ratio calc. |
|---|---|---|---|---|
| $\bar{1}$ 1 0 4 | 15.24 | 15.36 | 1.192 [a] | 1.198 |
| $\bar{1}$ 2 $\bar{1}$ 6 | 16.90 | 17.60 | 1.853 [b] | 1.817 |
| $\bar{1}$ 2 $\bar{1}$ 3 | 32.00 | 32.39 | 1.151 [c] | 1.141 |
| $\bar{1}$ 2 $\bar{1}$ 3 | " | " | 1.000 [d] | 1.000 |

Table 1: Comparison between experiment and calculations on the basis of a hexagonal Mg-type structure with lattice parameters a = 361 pm and c = 569 pm. First and second row: tilting angle of zone with respect to [0 0 0 1]. Third row: ratio of $R_{hkil}$ values of reflections spanning the diffraction patterns. Fourth row: corresponding inverse $d_{hkil}$ ratios. [a] 2 $\bar{2}$ 0 1 and 1 1 $\bar{2}$ 0 reflections; [b] $\bar{1}$ 2 1 $\bar{1}$ and 1 0 $\bar{1}$ 0 reflections; [c] 0 1 $\bar{1}$ $\bar{1}$ and 1 0 $\bar{1}$ 0 reflections; [d] 0 1 $\bar{1}$ $\bar{1}$ and 1 $\bar{1}$ 0 1 reflections.

We have carried out tilting diffraction experiments on a single grain to determine the symmetry and angular distance of zone axes around [0 0 0 1]. The results are summarized in Fig. 3, which displays the zone axes relevant for structure type determination within reach of the double-tilt goniometer stage. The [0 0 0 1] zone is surrounded by six <$\bar{1}$ 1 0 4> and six <$\bar{1}$ 2 $\bar{1}$ 6> zone axes (only half of which are shown).

The [$\bar{1}$ 1 0 4] zone is tilted by 15.24° with respect to [0 0 0 1] and shows a rectangular pattern that is repeated by five further zone axes rotated by 60° around [0 0 0 1]. The [$\bar{1}$ 2 $\bar{1}$ 6] zone is tilted by 17.60° with respect to [0 0 0 1] and shows a pattern with a different aspect ratio of the two reflections spanning the rectangular pattern. Five further equivalent zone axes are arranged around [0 0 0 1] rotated by angles of 60°. The dark lines represent the Kikuchi lines as seen in the TEM. The arrangement of the zone axes as well as the Kikuchi lines perfectly reproduces the hexagonal structure. A further zone axis displaying an almost hexagonal pattern, [$\bar{1}$ 2 $\bar{1}$ 3], which is tilted by 32° with respect to [0 0 0 1] was recorded. Note that some of the diffraction patterns contain very weak additional reflections, the origin of which is not clear. They may be due to double diffraction, minor deviations from perfect disorder, or electron diffraction taking place in neighboured grains. These reflections were not taken into account in our analyses.

All zone axes, Kikuchi lines and reflections can consistently be indexed on the basis of a hexagonal structure. We compared calculated and experimental tilting angles of different zone axes as well as distances of reflections from the direct beam $R_{hkil}$. For the latter, we compared the ratio of $R_{hkil}$ values of reflections spanning a diffraction pattern to the corresponding calculated $d_{hkil}$ ratio. Due to the relationship $R_{hkil}d_{hkil}$ = const, where the imprecisely known constant is given by the electron wavelength and the camera length of the microscope, these ratios should inversely correspond. Good agreement could be achieved using the lattice parameters a = 361 pm and c = 569 pm. These lattice parameters correspond to the weighted average of the parameters of the constituting elements. A comparison of the experimental and calculated quantities using these parameters is presented in Table 1.

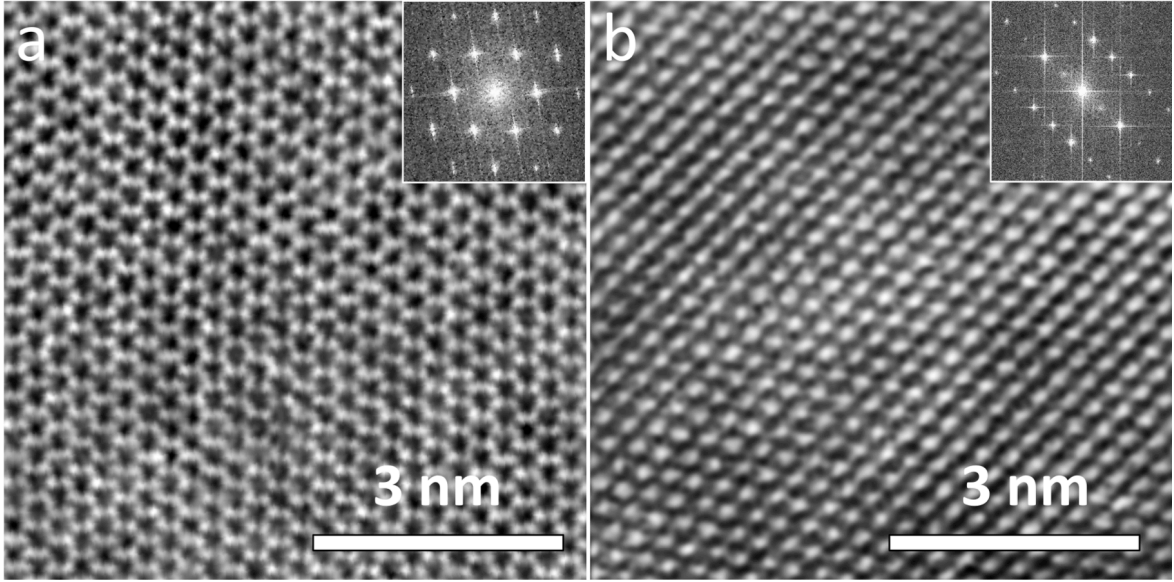

Figure 4. HAADF-STEM micrographs of equiatomic Ho-Dy-Y-Gd-Tb with insets displaying the corresponding FFTs. (a) along the [0 0 0 1] direction. (b) along the [1 1 $\bar{2}$ 0] direction.

Fig. 4 shows STEM micrographs, which were slightly FFT-filtered for high-frequency noise reduction. Fig 4(a) was taken along the [0 0 0 1] direction and in consistency with a Mg-type structure displays a pattern of interconnected hollow hexagons. The inset is a FFT of the image also reflecting the hexagonal structure. Fig. 4(b) was taken along [1 1 $\bar{2}$ 0] and the FFT in the inset corresponds to the diffraction pattern in Fig. 2(c). The image is also in consistency with a Mg-type structure. In particular, the atomic arrangement along this direction excludes a P6$_3$/*mcm* space group, which would lead to a pattern of paired atomic rows parallel to the [0 0 0 1] direction. This significantly differs from the almost hexagonal pattern expected for the P6$_3$/*mmc* space group, which directly corresponds to the experimentally observed HAADF-STEM image. This observation is consistent with our electron diffraction results. Both images, which were taken with a Z-contrast sensitive HAADF detector, show that also on this small scale the material is perfectly homogeneous. There is no indication of precipitation or ordering.

The lattice parameters directly measured from the image amount to a = (363 ± 30) pm and c = (566 ± 20) pm, which is in agreement with the weighted average of the parameters of the constituting elements. It was observed earlier that the a-lattice constant in cubic HEAs follows a rule of mixtures.[5] Our results indicate that a rule of mixtures also applies to the c-lattice parameter and the c/a ratio.

Our electron diffraction and STEM results are consistent with a Mg-type hexagonal structure. That is, the unit cell has space group P6$_3$/mmc (No. 194) and two atoms per unit cell, on the (1/3, 2/3, 1/4) position and its symmetric equivalent. The structure of the HEA is identical to the structures of the constituting elements, which are also of Mg-type.

Criteria for the formation of HEAs are commonly discussed in terms of radius differences and a parameter Ω referring to the thermodynamic properties of the alloy [6]. The radius difference $\delta r = \sqrt{\sum_i c_i (1 - r_i/\bar{r})^2}$, where $r_i$ are the element radii, $c_i$ the mole percentages, and $\bar{r}$ is the weighted average radius, should be as small as possible. A value below about 6.5 % for HEAs in general and below 3.8 % for single-phase HEAs [7] is required. In our case $\delta r$ = 0.77 %. The thermodynamic parameter $\Omega = T_m \Delta S_{mix} / |\Delta H_{mix}|$, where $T_m$ is the weighted average melting temperature, $\Delta S_{mix}$ the mixing entropy, and $\Delta H_{mix} = \sum_{i,j,i \neq j} c_i c_j H_{ij}$,[6] should exceed 1.1 (which means that the entropy contribution to the free energy should dominate the enthalpy).

Since the alloy investigated is equiatomic, $\Delta S_{mix}$ reaches the maximum value 1.61 R,[1] where R= 8.314 J K$^{-1}$ mol$^{-1}$ is the gas constant. $\Delta H_{mix}$ calculated using values from the tables provided by Takeuchi and Inoue [8] amounts to 0 kJ mol$^{-1}$. For Ho-Dy-Y-Gd-Tb not only the sum of the mutual mixing enthalpies is zero but each individual component $H_{ij}$ = 0 kJ mol$^{-1}$ for all $i$ and $j$. The alloy thus fulfills an even stronger formation criterion. Formally, taking the calculated values for $\Delta H_{mix}$ [8] into account, the parameter Ω in our case diverges.

Further criteria supporting HEA formation are equal crystal structures, a high mutual solubility in the binary phase diagrams of all constituting elements, and, from a practical point of view, close melting points. All formation criteria are ideally fulfilled for the system investigated.

Note that the same is true for numerous further RE elements. Similarly positive conditions are present for La, Er, Tm, Lu, Pr, Pm and Nd. If we exclude Pm due to its radioactivity, in total we are left with 11 practically useable elements for which we expect the formation of hexagonal HEAs. Hence HEAs with higher numbers of components up to 11 and numerous combinations should be possible. The number of possible equiatomic five-element HEAs is 462, and altogether 1486 different HEAs with five to 11 different elements can be combined. The elements Ho, Dy, Y, Gd, and Tb used for the alloy investigated in this work, which can be considered a feasibility study, were chosen more or less arbitrarily from the potential candidates due to their immediate availability in our stock. We should point out that some of the elements used undergo a high-temperature transition from hexagonal to cubic structure at 95 % (Gd and Tb) and 98 % (Dy and Y) of their melting point. The exclusive use of elements that have a hexagonal structure up to their melting point may lead to a larger grained structure. This however would limit the system to the four elements Er-Tm-Lu-Ho.

The list can probably be further extended by taking into account elements that have a different crystal structure but also similar radii and low mixing enthalpies with the elements mentioned above. Ce, for example, shows high solubility with all RE elements (with the presence of a miscibility gap, however) and mixing enthalpies between 0 and 1 kJ/mol. The same seems to hold for Sm, as far as phase diagrams are available.

We have chosen an equiatomic alloy composition for the present work but we expect that due to the similarity of the constituting elements and hence generous compliance of the formation criteria, the existence ranges of the HEAs are very wide. The production of single-phase ingots with different compositions should therefore also be possible. This expected tolerance to composition variation may open up the possibility to form single-phase HEAs using Mischmetall, which is a more or less undefined alloy of RE elements in various naturally occurring proportions. Since Mischmetall is considerably cheaper than high-purity single RE elements, this would eliminate the economical disadvantage of the RE-systems under consideration.

The hexagonal HEAs presented may offer a unique chance to study the magnetism of RE-systems. Mainly related to their 4f electron configurations, RE-elements possess highly complex magnetic behavior displaying a wide variety of properties [9]. The systems presented are unique in the sense that they are topologically ordered but possess random bonds. In the present HEA systems the included elements and the composition can most freely be varied at constant structure, which may allow for insightful comparative measurements. Furthermore, hexagonal HEAs including Tm, Yb, and Ce may provide a way to study heavy-fermion behavior in a disordered system.

The hexagonal high-entropy systems also include the possibility to incorporate intentional second phases, i.e. for the increase of strength, high-temperature applicability, etc. Potential candidates are Yb, Eu, Zr, and Hf, which have a high melting temperature and a high positive mixing enthalpy with the RE elements considered. The mixing enthalpies with these elements are between 6 and 18 kJ mol$^{-1}$, which will lead to segregation.

We should also point out that recently Gao and Alman [10] predicted the existence of a hexagonal HEA in the system Co-Os-Re-Ru. In this system the formation criteria are indeed well fulfilled and all constituting elements are mutually soluble. The drawback of this system is that Os reacts with oxygen at room temperature forming the highly toxic osmium tetroxide. Furthermore for the elements proposed, the melting points are strongly different (1768 K (Co) vs. 3459 K (Re)) making the growth procedure practically difficult. Further candidates along these lines are Co-Re-Ru-Tc or the five-element system Co-Re-Ru-Tc-Os, both of which involve the radioactive element Tc.

In conclusion, in this paper we report on the observation of a HEA in the HoDyYGdTb alloy system. Our samples are homogeneous and single phased, and of hexagonal Mg-type structure.

**Acknowledgments** The authors thank Ms D. Meertens for focused ion-beam preparation of the TEM samples, and W. Steurer, J. Dolinsek, and M. Heggen for inspiring discussions.

**References**

[1] Yeh JW, Chen SK, Lin SJ, Gan JY, Chin TS, Shun TT, Tsau CH, Chang SY. Nanostructured high-entropy alloys with multiple principal elements: novel alloy design concepts and outcomes. Adv. Eng. Mater. 2004;6:299 – 303.

[2] Cantor B, Chang ITH, Knight P, Vincent AJB. Microstructural development in equiatomic multicomponent alloys. Mat. Sci. Engng A 2004;375-377:213 - 218.

[3] Wei Q, Wanderka, N, Schubert-Bischoff P, Macht MP, Friedrich S. Crystallization phases of the Zr41Ti14Cu12.5Ni10Be22.5 alloy after slow solidification. J. Mater. Res. 2000;15:1729-1734.

[4] Morniroli JP, Steeds JW. Microdiffraction as a tool for crystal structure identification and determination. Ultramic. 1992;45:219-239.

[5] Senkov ON, Scott JM, Senkova SV, Miracle DB, Woodward CF. Refractory high-entropy alloys. Intermetallics 2010;18:1758-1765.

[6] Yang X, Zhang Y. Prediction of high-entropy stabilized solid-solution multicomponent alloys. Mat. Chem. Phys. 2012; 132: 233 – 238.

[7] Miracle DB, Miller JD, Senkov ON, Woodward C, Uchic MD, Tiley J. Exploration and development of high entropy alloys for structural applications. Entropy 2014;16:494-525.

[8] Takeuchi A, Inoue A. Classification of bulk metallic glasses by atomic size difference, heat of mixing and period of constituent elements and its application to characterization of the main alloying element. Mat. Trans. 2005;46: 2817 – 2829.

[9] Jensen J, Mackintosh AR. Rare earth magnetism: Structures and excitations. Oxford: Clarendon press; 1991.

[10] Gao MC, Alman DE. Searching for next single-phase high-entropy alloy compositions. Entropy 2013; 15: 4504 – 4519.